\documentclass{skads2009}
\usepackage{graphicx}
\begin{document}
   \title{SKA HI end2end simulation 
	}

   \author{H.-R. Kl\"ockner\inst{1}, R. Auld\inst{2}, I. Heywood\inst{1}, D. Obreschkow\inst{1}, F. Levrier\inst{3}\and
          S. Rawlings\inst{1} }

        \authorrunning{Kl\"ockner, Auld, Heywood, Obreschkow, Levrier, \& Rawlings }
   \institute{Oxford Astrophysics, Denys Wilkinson Building, Keble Road, Oxford, OX1 3RH, United Kingdom\thanks{This work was supported by the
   European Commission Framework Program 6, Project SKADS, Square
   Kilometre Array Design Studies (SKADS), contract no 011938.}
   \and
   Cardiff University, Queens Building, The Parade, Cardiff, CF24 3AA, United Kindom$^*$
   \and
   LERMA/LRA -UMR 8112 - Ecole Normale Sup\'erieure, 24 rue Lhomond, 75231 Paris CEDEX 05, France $^*$}

 \abstract{The current status of the HI simulation efforts is
   presented, in which a self consistent simulation path is described
   and basic equations to calculate array sensitivities are
   given. There is a summary of the SKA Design Study (SKADS) sky simulation and
   a method for implementing it into the array simulator is presented. A short
   overview of HI sensitivity requirements is discussed and expected
   results for a simulated HI survey are presented.}
   \maketitle
%
%________________________________________________________________
%
\section{Introduction}

One of the key-science goals of the SKA is to detect most of the
neutral hydrogen (HI; 1420.4~MHz in the rest frame) content of
galaxies out to cosmological redshifts of z $\sim$ 1 (for further
details see e.g. {\it Science with the Square Kilometre Array}
eds. Carilli and Rawlings or {\it Cosmology, Galaxy Formation and
  Astroparticle Physics on the Pathway to the SKA} eds. Kl\"ockner,
Rawlings, Jarvis and Taylor).  The technical parameters that determine
the performance of the SKA have been identified previously during the
SKADS program and are defined in the Benchmark scenario
(\cite{bench}). At the moment the SKA design includes three distinct
telescope technologies in order to cover the required frequency range,
i.e. between a few hundreds of MHz to 10$-$25~GHz.  The parameter
space one needs to cover to assess the performance of the
low-frequency sparse dipole array, the mid-frequency aperature array
(AA), or the high-frequency dish array is enormous, and this is
impossible to accomplish via a single telescope simulation. To get a
basic understanding of the array parameters and their influence on the
quality and sensitivity of the final images, one can parameterise each
of the different antenna types in terms of a dish equivalent. For a
dish array, two basic parameters are the system equivalent flux
density (SEFD) and the image sensitivity ($\Delta$I). Respectively,
these are given by

\begin{equation}
{\rm SEFD} = \frac{{\rm T}_{\rm sys}  2  {\rm k}}{\eta_a \, {\rm A}},
\end{equation}

\noindent where T$_{\rm sys}$ is the system temperature [K], k is the Boltzmann constant, A is
the collecting area [m$^2$], and $\eta_a$ is the aperture efficiency, and (for a naturally weighted image) the image sensitivity can be calculated via:

\begin{equation}
{\rm \Delta I} = \frac{\rm SEFD}{\sqrt{{\rm N  (N-1) {\rm N_{\rm Stokes}} t} \Delta \nu  }},
\label{eq:sens}
\end{equation} 

\noindent where N is the number of Antennas, N$_{\rm Stokes}$ is the number of Stokes parameters, 
t integration time [s], and $\Delta \nu$ is the bandwidth [Hz] (\cite{ww1999}; a sensitivity 
calculator can be found at
www-astro.physics.ox.ac.uk/$\sim$hrk/ARRAY$_{-}$EXPOSURE.html).

In order to increase the image sensitivity both equations dictate that
the system temperature and the effective area are crucial to the
telescope performance. However in the case of continuum emission that
is spectrally well-behaved (spectral index of around zero), the image
sensitivity can be increased by trading the SEFD for increased
bandwidth. For spectral line observations, where the bandwidth is
tailored to the width of the expected signal, this cannot be done.
Bandpass stability also plays a crucial role in the image quality
in this regime.

In this article we describe an end-to-end (e2e) HI simulation plan
that is focused on exploring a more manageable parameter space defined
by the system temperature, effective area, and the spatial
configuration of the array.

%__________________________________________________________________

\section{Simulation}

Any aperture synthesis telescope acts as a spatial frequency filter
and an analytic determination of the final image quality is in most
cases impossible. Therefore the purpose of any interferometry
simulation is to define a practical sensitivity limit with respect to
the theoretical estimates. Generally one would like to have a full
array simulation that simulates the complete signal path from the
astronomical source up to and including the receiver electronics. Such
an approach is very complicated, computationally expensive and for the
purposes of a full SKA simulation completely impractical. The e2e
simulation can be broken down into individual components each of which
is treated as a standalone simulation.  If one is interested in the
electromagnetic properties (e.g.~the directional gain and its
stability) of individual dish designs one needs to take into account
the telescope structure and make use of a full electromagnetic
simulation (e.g. \cite{holler}). Similar simulations are needed if one
is interested in the performance of an aperature array and its primary
beam pattern (e.g. OSKAR; www.oerc.ox.ac.uk/research/oskar). So far
little has been done to simulated the electronic path from the
receiver to the correlator, but that is certainly in the scope of the
PrepSKA (www.jb.man.ac.uk/prepska) program. Finally, after the
correlation stage the data quality and array performance can be
evaluated by the simulating the response of individual baselines, or
by simulating and assessing the final image.

Radio astronomy software packages which are currently available and
offer simulation functionality (e.g. AIPS, CASA, MeqTrees) generally
focus on generating a model visibility set given a model sky and a set
of parameters which describe the observation. Such functionality has
emerged naturally, due to the self-calibration process relying on the
generation of model visibilities, however the sophistication of the
simulations offered by packages has generally evolved beyond this
fundamental role, particularly in the case of MeqTrees.

\begin{figure}
  \centering
  \includegraphics[bb=181 30 680 560,width=8.5cm,clip]{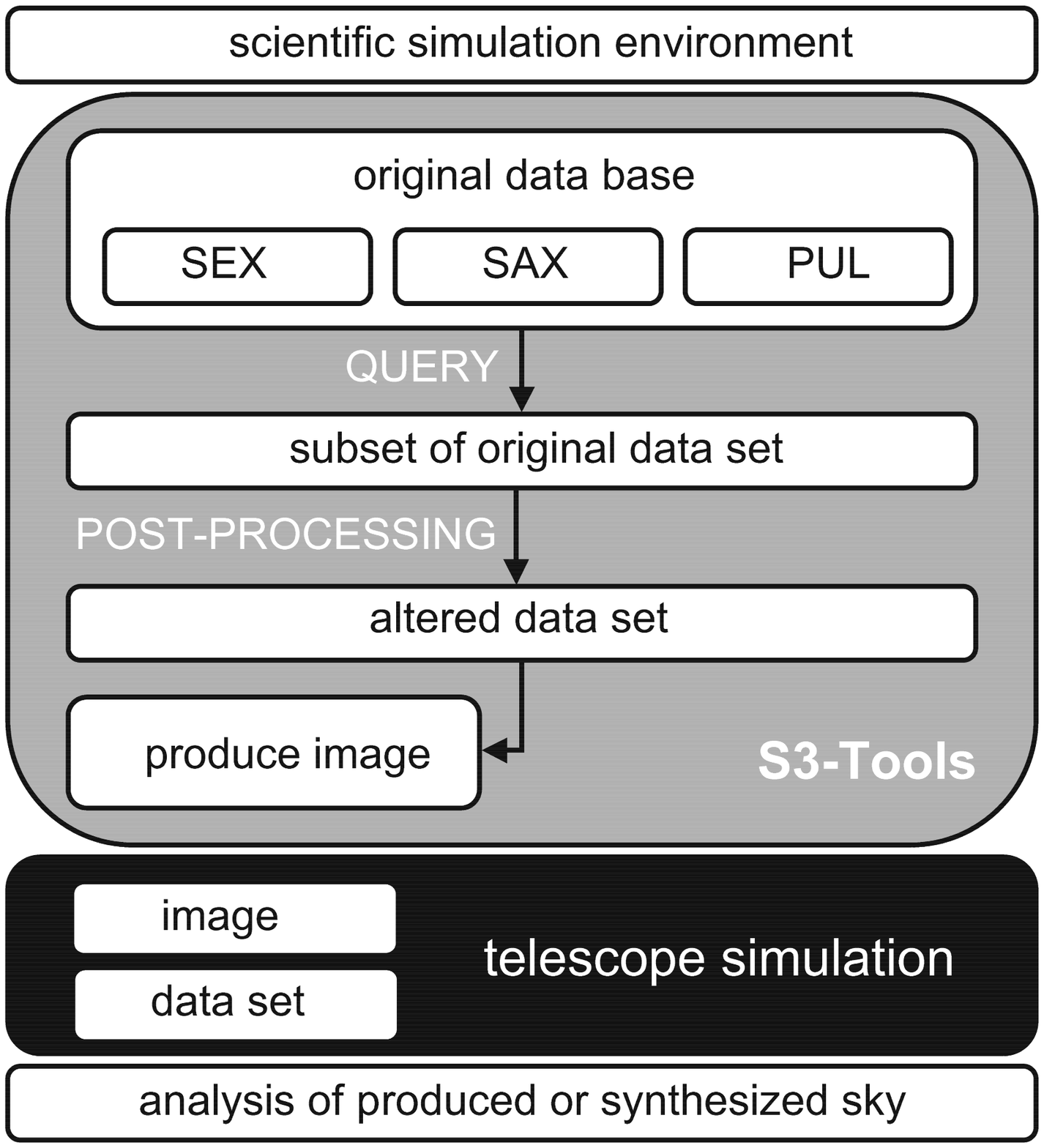}
  \caption{Overview of a full simulation path.
  }
  \label{FIG:simflow}
\vspace{-0.5cm}
\end{figure}

An overview of the developed e2e simulation path is shown in
Fig.~1. The grey box shows the steps we take to generate a model
sky. In our case this is based on the SKADS Simulated Skies (S$^3$),
which are a series of databases describing the properties of a range
of simulated astrophysical objects. The databases themselves are
discussed further in the next section.  Specific subsets of these
databases can be retrieved and processed, and finally converted into
either a 2-D image or a 3-D datacube (the third axis being frequency)
with optional Gaussian noise\footnote{A suite of Python-based routines
  with user-friendly GUIs, collectively known as the S3-Tools allows
  access, manipulation and imaging of the SKADS Simulated Skies.
  These tools have been developed by F. Levrier and a general overview
  can be found in this volume (Levrier, 2009 in these
  proceedings).}. These represent idealised radio skies which can then
be fed into a telescope simulation package. The array simulator is
represented in Fig.~1 by the black box, which will take the model sky
image and generate a model data set based on this. The final step is
the analysis of this end product. In the case of the e2e HI simulation
the analysis step involves running both the `idealised' and `observed'
HI datacubes through a source finder algorithm and comparing the two
resulting catalogues. This comparison can then be used to benchmark a
specific telescope design or observing strategy. In addition, the
ability of the S$^3$-tools to produce maps with Gaussian noise allows
to test the source finding algorithm and provide measures for
completeness studies.

\subsection{The sky simulation}

Here we present a short summary of the sky simulations that can be
used within the array simulator. Further details on the simulations
together with a webform which can be used to query the databases is
available on the Oxford S$^3$
webpage\footnote{s-cubed.physics.ox.ac.uk}. The full suite of
simulations is presented using two distinct products. Properties of
individual extragalactic objects and of Galactic pulsars are stored in
databases (SEX, SAX, PUL), whereas morphologically complex structures
such as the Global Sky Model (GSM) and the signals of the Epoch of
Reionization (EOR) are available as images.

The GSM is the radio foreground of our Galaxy which has
been modeled by the radio and (sub)- millimeter emission between
10~MHz to 100~GHz. In addition to the diffuse Galactic emission it also includes
emission from individual point sources e.g. supernova remnants
(\cite{gsm}; space.mit.edu/home/angelica/gsm).

The EOR images display the HI line signal of the Intergalactic medium
(IGM) during the Epoch of Reionization. This simulation covers the
redshift range between z = 5.6 and z = 23.6. In addition to the
ionization field, the effect of inhomogeneous heating of the IGM by
X-rays and the Lyman-$\alpha$ radiation field are taken into
account. The simulations have been produced in a cubic simulation box
with a side length of s$_{box}$ = 100 Mpc/h and a particle mass resolution
of 3$\times$10$^6$/h solar masses (Santos et al. 2008).

The semi-empirical simulation of extragalactic sources (SEX) describes
the radio continuum emission in a sky area of 20$\times$20 deg$^2$
out to a cosmological redshift of z = 20. As the name suggests, the sources were drawn from
observed (or extrapolated) luminosity functions and grafted onto an
underlying dark matter density field with biases which reflect their
measured large-scale clustering. This approach puts an emphasis on
modelling the large-scale cosmological distribution of radio sources
rather than the internal structure of individual galaxies. \\

\noindent Five source types of radio sources have been included in the simulation:
\begin{itemize}
\item[-]Radio-quiet AGN [1 core; 36,132,566 sources]
\item[-]Radio-loud AGN of the FRI class [1 core + 2 lobes; 23,853,132 sources]
\item[-]Radio-loud AGN of the FRII class [1 core + 2 lobes + 2 hot-spots; 2,345 sources]
\item[-]Quiescent star-forming galaxies [1 disk; 207,814,522 sources]
\item[-]Starbursting galaxies [1 disk; 7,267,382 sources]
\end{itemize}

\noindent For each of the source types, the database provides the
radio fluxes at observer frequencies at 151 MHz, 610 MHz, 1.4 GHz,
4.86 GHz and 18 GHz, down to flux density limits of
10~nJy. Intermediate frequencies can be determined by using the
S3-tools. In addition to the continuum emission this simulation
provides a rough estimate of the HI mass of the starbursting and
star-forming galaxies (\cite{wilman}). A secound version of these
simulations extending the simulated properties into the far-infrared
(principally for comparison with data from the Herschel satellite) has
now been completed (\cite{wilman2}).

The semi-analytic simulation (SAX) provides the properties of neutral
atomic (HI) and molecular (H$_2$) hydrogen in galaxies and associated
radio and sub-millimeter emission lines, i.e. the HI-line and various
CO transition lines. This simulation relies on the Millennium
simulation of cosmic structure (\cite{springel}), which reliably recovers comoving
length scales from 10 kpc to several hundred Mpc and galaxies with
cold hydrogen masses (HI+H2) above 10$^8$~M$_{\odot}$. 

There are two versions of the SAX database, reflecting the different
versions of the Millennium simulation: the full Millennium simulation
(s$_{box}$ = 500/h Mpc; $\sim$ 685 Mpc) and the smaller test version,
called the Milli-Millennium simulation (s$_{box}$ = 62.5/h Mpc; $\sim$
85.6 Mpc) where s$_{box}$ defines the diameter of the simulated box.
Both of the SAX databases have been produced by constructing a mock
observing cone from the corresponding simulation box. The opening
angle or the field of view (FoV) of these simulations therefore depends
on the values of the maximum redshift one requires (z$_{max}$), e.g. for a
redshift of 1 the FoV of the simulation would be 12$\times$12
deg$^2$. More information about the FoV of the simulation can be
obtained via the simulation page. Currently, radio continuum data is
not available, although efforts are being made to add this information
(Obreschkow et al. 2009a; Obreschkow et al. 2009b; Obreschkow et
al. 2009c)

S3-Tools is used to build mock radio maps or cubes in which the
radio emission of the extragalactic radio sources could be combined with
the diffuse radio emission of the GSM or EOR.

\subsection{Array simulation}

The developed array simulator\footnote{A copy of this simulator can be
  downloaded via www-astro.physics.ox.ac.uk/$\sim$hrk} is based on ``classical'' AIPS and
ParselTongue (\cite{aips,PT2006}), the Python interface to AIPS. The
core function of the array simulator makes use of the AIPS task {\tt
  UVCON}. The basic input of this task is a list of antenna locations,
together with properties of each antenna, such as diameter, system
temperatures and aperture efficiencies.  Additional inputs are the
total time of observation, the integration time per visibility and the
input sky model which also defines the pointing position on the
sky. The output is a standard UV-FITS data file in which the
visibilities correspond to the input model with added Gaussian noise
appropriate for the specified antenna characteristics. Dirty or
deconvolved images can be produced by invoking the task {\tt IMAGR}.
If the input model consists of a cube instead of a 2-D image then two
different simulation paths can be used to generate the ``observed
sky''. If the output map is to be a continuum image formed at the
central observing frequency, the visibilities are produced per
frequency step and are finally merged into a single visibility set. If
instead the output should be a 3-D datacube, the simulation produces a
unique visibility set at each frequency step, each of which is then
imaged independently. Each plane is finally combined into a cube by
using the task {\tt MCUBE}. For cases where the duration of the
observation equals or exceeds that necessary for a full UV coverage, the
visibilities are generated for a full synthesis and the noise scaled
down accordingly. This cuts down on both processing overheads and the size
of the resulting UV data files.

This style of simulation is well suited to the investigation of the
completeness of surveys as well as to the understanding of the imaging
capabilities when observing individual galaxies. In particular one can
investigate the quality of snapshot observations for different array
layouts.  There are, however, several AIPS--based limitations which
have a direct influence on the questions we can ask, and more
complicated simulations are needed to address these. Currently the
number of array elements which AIPS can handle is limited to 255
(E. Greisen 2009, private communication), which is sufficient for most of
the current arrays but is not enough for a full simulation of, for
example, the LOFAR array at the level of individual dipole
elements. One can circumvent this to some extent by combining several
elements into a single station.  This is relevant since LOFAR does
indeed perform correlations between the beamformed data from each
station, and not between individual dipoles. This is a model which the
SKA is likely to follow in order to minimize the data stream.

Two properties of a real interferometric array which are absent from
this simulation framework are primary beam effects and bandpass
variability. These are two factors which significantly affect the
dynamic range and image quality of a real observation, particularly
the former in the case of an aperture array.

Despite the various limitations we feel that AIPS is likely to be the
most tested software package and our approach provides a basic
simulation pipeline which can provide a reliable check for more
complicated simulation efforts (e.g. within MeqTrees or CASA).

In the redshift range around z~=~1 the SKA's mapping speed has the
potential to be revolutionized by mid-frequency aperture arrays (AA).
As mentioned before, the simulation software is limited to 255
antennas and it is necessary to combine the aperture arrays
into individual stations. The following equation shows how to
calculate the total area of an aperture array, assuming Hertzian
dipoles, and therefore the equivalent dish diameter of an AA station:

\begin{equation}
\rm A_{\rm AA} = N_{\rm tile}  N_{\rm dipole}  \frac{3}{8 \pi} (\frac{c}{\nu})^2 N_{\rm stokes}, 
\end{equation}
\noindent where N$_{\rm tile}$ is the number of tiles and N$_{\rm
  dipole}$ is the number of dipoles per tile.  Assuming a square
kilometre of collecting area the diameter of a SKA AA station would be
56~m. Note that an aperture efficiency of 80\% and a total of 255
stations has been assumed for a S$^3$ default SKA realization
(S$^3_{\rm real}$). This value only corresponds to observations
towards the zenith, and the effective station area varies with
elevation. However simulations to investigate the performance of such
a simplified AA are still valuable because the layout of the array
will directly affect the synthesized beam and therefore the imaging
capability.  Figure~3 shows a cut through the dirty beam pattern
showing a high sidelobe pattern which can be minimized by varying the
station distribution within the array configuration (e.g. AntConfig
can be used for this purpose; www.kat.ac.za/public/wiki/AntConfig).

The current design of the aperture array has two layouts that have the
same core, but differ in the outer regions (R. Bolton 2009, private
communication). The core has a radius of 2.5~km and has 165 randomly
placed stations which are separated at least by 96~m. 

\begin{itemize}
\item{The ``concentrated'' layout has 72 stations beyond the core
    placed in 5 spiral arms out to 10 km (radius). Beyond this 13
    stations are placed on the same spiral arms out to 180~km (S$^3_{\rm real}$C).}
\item{The ``not-concentrated'' layout has 85 stations beyond the core
    logarithmically placed in 5 spiral arms out to 180~km (S$^3_{\rm real}$NC).}
\end{itemize}

\noindent Figure~2 shows the UV coverage of a test simulation of 1
hour duration. For displaying purposes each visibility has an
integration time of 10~minutes. The high density of visibilities and
the filled central core will provide high sensitivity and image
fidelity when observing diffuse HI emission. However deconvolution of
such structure will be difficult because of the dirty beam pattern
shown in Fig.~3. The broad sidelobe pattern will add ambiguities
during attempts to recover diffuse, extended HI emission.

\begin{figure}
  \centering
  \includegraphics[width=8cm,clip]{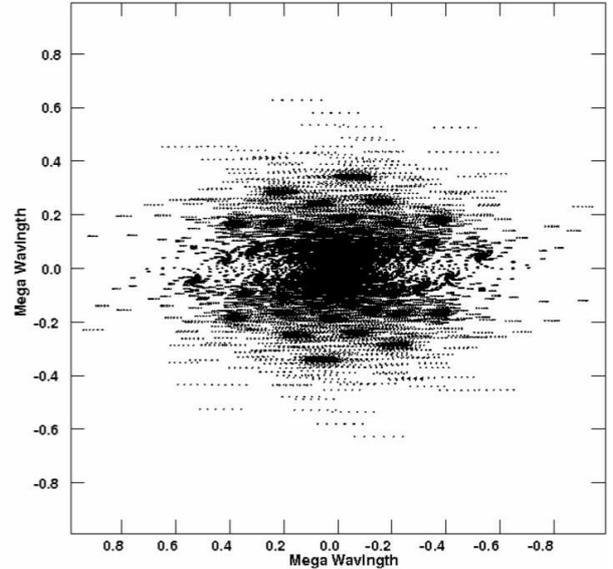}
  \caption{The UV coverage of the ``concentrated'' aperture-array
    layout (S$^3_{\rm real}$C). The simulation is based on 1~hr integration with an
    integration time of 10 minutes per visibility. Note that the 10
    minutes integration per visibility is for displaying purposes
    only.}
  \label{FIG:uvcover}
\end{figure}

\begin{figure}
  \centering
  \includegraphics[bb=40 140 565 642,width=8cm,clip]{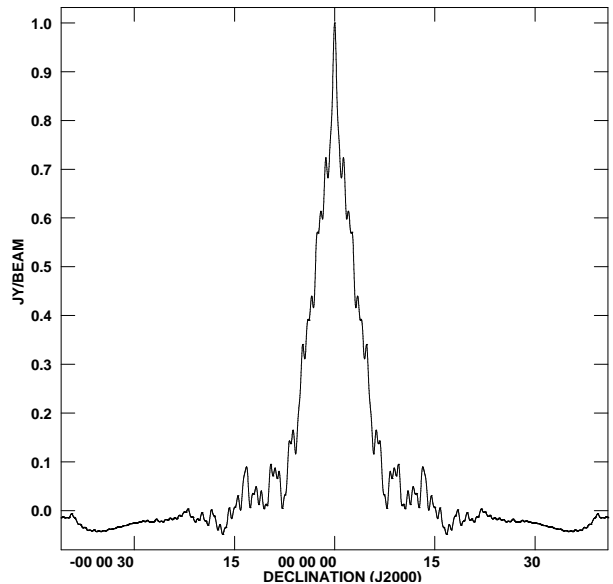}

  \caption{Cut through the synthesized beam (S$^3_{\rm real}$C). Due to the high density
    of stations in the core area, the synthesizes beam profile
    plateaus around 30 arcsec and shows strong sidelobes at a level of
    $\sim$10\% .  }
  \label{FIG:beamcut}
\end{figure}

A final limitation of the simulation pipeline arises at the imaging stage. 
The images that can be handled by AIPS are limited to 8096 pixels per
dimension. For the current SKA layout the maximum baseline length of 
360~km would provide sub-arcsec angular resolution at frequencies
higher than 200~MHz. Such resolution limits the field-of-view that
can be simulated, and producing a simulation covering several square
degrees requires the sky to be divided into sub-patches. For example,
the spatial resolution of the array configuration at 700~MHz is
0.3~arcsec. If we Nyquist-sample the sky then these images
need to have a pixel resolution of about 0.1~arcsec per pixel. Taking
the AIPS limitation into account, such images would cover a sky area of
0.67$\times$0.67 deg$^2$ and thus to simulate a 4~deg$^2$ field 9
sub-images are necessary.

\subsection{The HI e2e simulation}

\begin{figure}
  \centering
  \includegraphics[bb=30 180 540 580,width=8.5cm,clip]{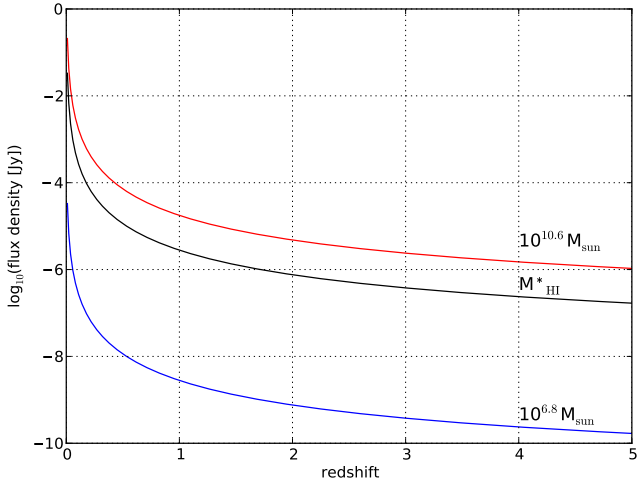}
  \caption{Flux density of HI masses with a fixed line width of
    164~kms$^{-1}$ versus redshift. The three lines show the
    entire HI mass range of the HIPASS survey. Note for detection
    arguments one needs to assume a channel resolution similar to the
    HI line width, otherwise the sensitivity is reduced via
    Eq.~\ref{eq:sens}.}
  \label{FIG:sensi}
\end{figure}

The anticipated HI simulation will make use of the SAX simulated
sky. These simulations do not include a physical model of the
associated continuum emission. However, to evaluate the influence of
continuum emission to the HI simulation a mock continuum
component may be added to the line emission (using the task {\tt
UVMOD}). The HI emission of each galaxy has been pasted into the
model cube by using S3-tools, selecting the ``Oxford'' HI templates in the map making tool. The
following equation is given to provide a general understanding of the detectability of HI. 
The basic relationship between HI mass and HI
line flux density is defined by

\begin{equation}
  M_{\rm HI} = \frac{2.36 \times 10^5}{(z + 1)}  {\rm D^2_L}  \frac{\pi}{\sqrt{\rm 2 ln(2)}} {\rm S_p} \Delta {\rm V} ,
\end{equation}

\noindent where D$_{\rm L}$ is the luminosity distance [Mpc], assuming
a Gaussian line profile where $S_{\rm p}$ is the peak flux density
[Jy] and $\Delta {\rm V}$ is the full line width measured at
half-maximum [FWHM; kms$^{-1}$]. To investigate how much of the HI
mass function one can trace with the SKA, the expected HI flux density
is shown in Fig.~4. The HI flux density has been calculated by using
average values over the HIPASS sample for the HI line width (FWHM; 164
kms$^{-1}$) and for the HI masses (range between 2$\cdot$10$^{7}$ ---
6$\cdot$10$^{10}$~M$_{\odot}$) (\cite{zwaan1,zwaan2}). Note that in Fig.~4
to detect the HI mass at specific redshift one assumes that the entire
HI signal is confined within one channel. For a freely chosen 1-sigma
rms noise of 1~$\mu$Jy one could detect M$^*$$_{\rm HI}$ galaxies at
the 3-sigma level out to a redshift of 1 (Note that a rms noise of
1~$\mu$Jy would correspond to an integration time of 36 hours, assuming
a channel width of 250~kHz and a T$_{\rm sys}$ of 50~K.). This is the
anticipated aim of the SKA and the HI simulations need to be able to
investigate such a cosmic volume.  Furthermore, if there is little
evolution in the HI mass function one could expect to detect the most
massive HI galaxies up to redshifts of about 4.

The FoV of the Millenium simulation box corresponds to an area of
116$\times$116~deg$^2$ at redshift 0.1, which shrinks to
5.4$\times$5.4~deg$^2$ at redshift 4. At low redshifts a HI
simulation would help to investigate potential systematic errors in a
measurement of the faint end of the HI mass function. Such a
simulation would require around 30276 individual simulation runs,
based on the field-of-view limitations per-facet, as discussed above.
This is a computationally very expensive exercise, and is only
worthwhile once the SKA array configuration has been finalised.  For
investigating and optimising array configurations it makes more sense
to perform smaller scale simulations which still allow us to quantify
the performance of the array.

%__________________________________________________ One column table
   \begin{table}
     \caption[]{Simulation parameters at various redshifts. Bandwidth corresponding to a fixed redshift interval of $\Delta z = 0.1$. The fraction of the HIPASS volume to the co-moving volume of 1 square degree and $\Delta z = 0.1$ (e.g. at redshift 1 such volume would correspond to 0.0008~Gpc$^3$ deg$^{-2}$). The FoV of the SAX sky simulations (values are obtained from s-cubed.physics.ox.ac.uk). Expected HI sources per square degree
       for a flux limit of 3~$\mu$Jy (assuming a rms of 1~$\mu$Jy). For example to investigate the HIPASS volume at redshift 1 a 4$\times$4 deg$^{2}$ field must be simulated and need to be split into 36 individual sub-array-simulations to handle the AIPS image limitation.}
         \label{tab:1}
     $$ 
         \begin{array}{ccccc}
            \hline
            \noalign{\smallskip}
            {\rm redshift}      &  {\rm bandwidth} & \frac{\rm HIPASS vol}{\rm co-volume} & {\rm SAX sky} & {\rm SAX sources}\\

                  &  {\rm [MHz]} & {\rm [deg^{2}]} & {\rm [deg^2]} & {\rm [deg^{-2}]}\\

            \noalign{\smallskip}
            \hline
            \noalign{\smallskip}
            0.5 & 59 & 37 & 21.4 \times 21.4 & 15136\\
            1 & 34  & 16 & 12.0 \times 12.0 & 7136\\
            1.5 & 22 & 12 & 9 \times 9  & 1897 \\
            2 & 15 & 11 & 7.5 \times 7.5 & 816 \\
            \noalign{\smallskip}
            \hline
         \end{array}
     $$ 
\vspace{-0.2cm}
\end{table}

We propose such a HI simulation here, which will partly match the
co-moving volume of the HIPASS dataset. This requires a smaller number
of individual simulation runs, and it is thus possible to re-run the
simulation several times.

HIPASS is a blind HI radio survey of the sky at declinations
southwards of 25 degrees. However to calculate the co-moving volume we
use the initially presented catalogue covering the hole southern
hemisphere. The survey covers a sky area of 21314~deg$^2$ , a redshift
range of z = 0.001--0.042 (corresponding to a co-moving volume of
0.013 Gpc$^3$), and has a sensitivity limit approximately
corresponding to a peak flux density of 0.05~Jy (see \cite{zwaan1} for
a full description of the sample completeness). This survey yields a
total number of 4315 sources, all of which are subsequently identified
with optical galaxies (\cite{doyle}).

Using the SAX simulation to mimic the HIPASS survey no
extrapolation of the sky simulation is needed because the simulated
volume is large enough to cover the entire HIPASS volume. In fact this
can been shown by comparing the HIPASS co-moving distance of 171~Mpc
(z=0.042 and using the cosmological parameter of the Millenium
simulation) which is smaller than the radius of the simulation box,
i.e. 500/2~Mpc/h $\sim$ 342~Mpc.

Using the online query of the SAX simulation and defining the redshift
and the sensitivity specifications\footnote{SAX query input: select
  count(*), from galaxies$_{-}$line where zapparent between 0.001 and 0.042
  and hiintflux*hilumpeak$>$0.05 } of the HIPASS catalogue one obtain
4545 sources. This means that the SAX simulation predicts that the
HIPASS catalog contains 4545 sources, which matches the observed
number of 4315 sources within 5~\%. This difference can be partially
attributed to the fact that HIPASS yields a continuously varying
completeness function rather than a strict peak flux limit. Assuming
no flux limits the SAX simulation predicts 772120 source within
the HIPASS volume, i.e. 180-times more sources than picked up by HIPASS.
However Table~\ref{tab:1} shows the expected number counts assuming a
minimum peak flux density of 3~$\mu$Jy. The simulation still contain
enough source to address e.g. the study of the faint end of the HI
mass function and to make statistical significant predictions out to
high cosmological distances.

\begin{figure}
  \centering
  \includegraphics[bb=70 195 526 605,width=8.7cm,clip]{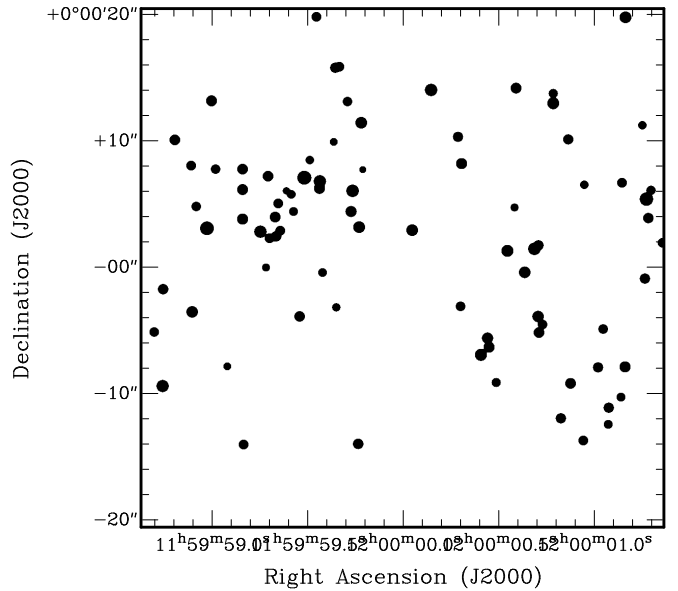}
  \caption{Example of an HI input sky. The cube has 11 spectral
    channels covering a frequency range of 11$\times$62.5~kHz. For
    illustration purposes the image shows the channel averaged line
    emission (using {\tt SQASH}) and in addition the averaged image has been
    convolved with a 0.3~arcsec Gaussian beam (using {\tt CONVL}).}
  \label{FIG:mom0}
\end{figure}

Figure~5 shows the HI intensity map of an input cube. No
continuum emission has been added and the individual galaxies are
unresolved. The corresponding ``observed sky'' is shown in Fig.~6. In
order to analyse these images the automated source finder,
Duchamp\footnote{www.atnf.csiro.au/people/Matthew.Whiting/Duchamp}
will be used to generate ``observed'' source catalogues for comparison
to the input catalogues. An important part of this analysis will be to
produce idealised input images with purely Gaussian noise of an
equivalent level to cross--check the reliability and completeness of
the source finding software in the presence of the image artifacts
introduced by an interferometer.

\begin{figure}
  \centering
  \includegraphics[bb=30 150 540 645,width=8cm,clip]{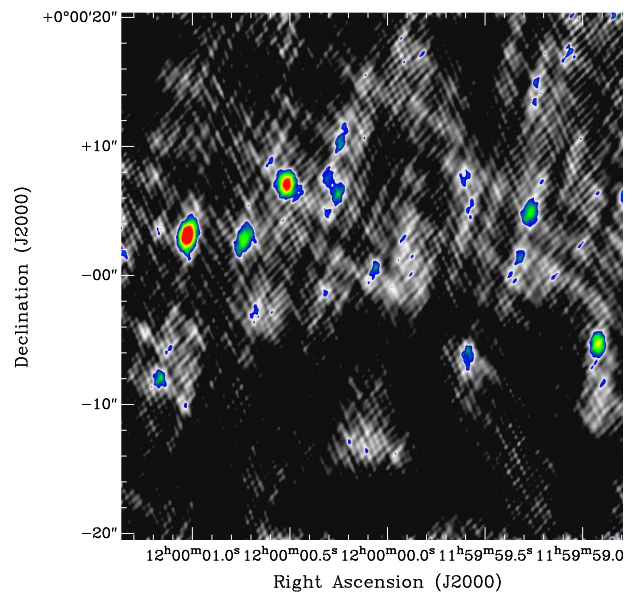}
  \caption{Example of a simulated HI sky. The rms of this image
    including sources is of the order of 8~$\mu$Jy. The cleaned image
    (using 200 clean components) displays the channel averaged line
    emission.\vspace{0cm}}
  \label{FIG:mom0sim}
\end{figure}

\section{Conclusions}

A full e2e simulation path has been developed. The array simulator
makes use of images or datacubes based on the S$^3$ catalogues to
simulate an observation. In the simulations no errors due to
calibration or telescope hardware have been introduced, but a
simplistic treatment of gain and phase errors could be included at
later stages.  The analysis of the simulated data sets (either images
or cubes) will make use of automated source finder software. In
addition to this, a more sophisticated analysis of the simulated data
is possible because the simulator produces visibility data as well as
images. For example one could investigate the sensitivity of an array
to diffuse emission by using the UV-gap ($\frac{\Delta{\rm U}}{\rm
  U}$) analysis techniques (\cite{lalduu}) or by analysing the
statistics of Fourier phases (\cite{francois}).

The proposed HI simulations match the requirements for studying the
capability of the SKA aperture array when imaging HI structures in
nearby galaxies. Such simulations will also investigate the impact of
different telescope designs on the proposed SKA blind HI surveys.  The
input HI sky will have an equivalent co-moving volume to that of the
HIPASS survey, and this relatively small volume makes it feasible to
re-run the simulations with different parameters within a reasonable
time. With this setup we are able to analyse the ``observed skies''
and study the influence that the antenna layout has on a blind HI
galaxy survey by investigating following points:

\begin{itemize} 
\item[-] completeness (peak flux density)
\item[-] positional accuracy
\item[-] redshift determination
\item[-] necessity of subtracting continuum emission and our ability to do so
\end{itemize} 

\begin{acknowledgements}
HRK would like to thank Rosie Bolton for the two array configuration files.
\end{acknowledgements}

\end{document}